# Gut decisions based on the liver: A radiomics approach to boost colorectal cancer screening

Anna Hinterberger[1,2*], Jonas Bohn[3,4,5,6*], Dasha Trofimova[3,7], Nicolas Knabe[8], Julia Dettling[8], Tobias Norajitra[3,4,9], Fabian Isensee[3,7], Johannes Betge[1,10,11,12], Stefan O. Schönberg[8], Dominik Nörenberg[8], Sergio Grosu[13], Sonja Loges[1,14,15], Ralf Floca[3,6,9], Jakob Nikolas Kather[16,17,18], Klaus Maier-Hein[3,4,6,7,9,19,20*], and Freba Grawe[1,2,8*]

[*] Equal contribution
[1]DKFZ Hector Cancer Institute at the University Medical Center Mannheim, Germany.
[2]Junior Clinical Cooperation Unit Translational Molecular Imaging in Oncologic Therapy Monitoring (E310), German Cancer Research Center, Heidelberg, Germany,
[3]Division of Medical Image Computing, German Cancer Research Center (DKFZ), Heidelberg, Germany
[4]Translational Lung Research Center (TLRC), Member of the German Center for Lung Research (DZL), Heidelberg, Germany
[5]Faculty of Biosciences, Heidelberg University, Heidelberg, Germany
[6]National Center for Tumor Diseases (NCT Heidelberg), Heidelberg, Germany.
[7]Helmholtz Imaging, Heidelberg, Germany
[8]Department of Radiology and Nuclear Medicine, University Medical Center Mannheim, Heidelberg University, Mannheim, Germany.
[9]Pattern Analysis and Learning Group, Heidelberg University Hospital, Heidelberg, Germany
[10]Department of Medicine II, University Medical Center Mannheim, Medical Faculty Mannheim, Mannheim, Germany
[11]Junior Clinical Cooperation Unit Translational Gastrointestinal Oncology and Preclinical Models, German Cancer Research Center, Heidelberg, Germany
[12]German Cancer Consortium, DKTK, Heidelberg, Germany
[13]Department of Radiology, University Hospital, LMU Munich, Munich, Germany
[14]Division of Personalized Medical Oncology (A420), German Cancer Research Center (DKFZ), Heidelberg, Germany.
[15]Department of Personalized Oncology, University Hospital Mannheim, Medical Faculty Mannheim, University of Heidelberg, Mannheim, Germany.
[16]Else Kroener Fresenius Center for Digital Health, Faculty of Medicine and University Hospital Carl Gustav Carus, TUD Dresden University of Technology, Dresden 01307, Germany.
[17]Department of Medicine I, University Hospital Dresden, Dresden, Germany.
[18]Medical Oncology, National Center for Tumor Diseases (NCT), University Hospital Heidelberg, Heidelberg, Germany.
[19]Faculty of Medicine, University of Heidelberg, Heidelberg, Germany
[20]Faculty of Mathematics and Computer Science, Heidelberg University, Heidelberg, Germany

**Abstract**

Background and Aims:

Non-invasive colorectal cancer (CRC) screening represents a key opportunity to improve colonoscopy participation rates and reduce CRC mortality. This study explores the potential of the gut–liver axis for predicting colorectal neoplasia through liver-derived radiomic features extracted from routine CT images as a novel opportunistic screening approach.

Methods:

In this retrospective study, we analyzed data of 1,997 patients who underwent colonoscopy and abdominal CT. Patients either had no colorectal neoplasia (n=1,189) or colorectal neoplasia ($n_{total}$=808; adenomas n=423, CRC n=385). Radiomic features were extracted from 3D liver segmentations using the Radiomics Processing ToolKit (RPTK), which performed feature extraction, filtering, and classification. The dataset was split into training (n=1,397) and test (n=600) cohorts. Five machine learning models were trained with 5-fold cross-validation on the 20 most informative features and the best model ensemble get selected based on the val AUROC

Results:

The best radiomics-based XGBoost model achieved a test AUROC of 0.810 [95% CI: 0.767–0.837], clearly outperforming the best clinical-only model (test AUROC: 0.457 [95% CI: 0.411–0.506]). After Youden index-based threshold optimization, the final model reached a test sensitivity of 74.1% and specificity of 72.3% for predicting the presence of colorectal neoplasia. Subclassification between colorectal cancer and adenoma showed lower accuracy (test AUROC: 0.674 [95% CI: 0.606–0.741]).

Conclusions:

Our findings establish proof-of-concept that liver-derived radiomics from routine abdominal CT can predict colorectal neoplasia. Beyond offering a pragmatic, widely accessible adjunct to CRC screening, this approach highlights the gut–liver axis as a novel biomarker source for opportunistic screening and sparks new mechanistic hypotheses for future translational research.

# Introduction

Colorectal cancer (CRC) is largely preventable through the detection and removal of precancerous lesions during screening colonoscopy [1,2]. Despite the availability of organized screening programs, CRC remained the third leading cause of cancer-related mortality in Europe in 2022, likely reflecting low screening participation rates [3,4], which vary considerably, with one German study even reporting rates as low as 20% [5–7]. Given the recommended participation rate of 80% to achieve a substantial reduction in CRC mortality [4], current levels are clearly insufficient, underscoring the urgent need to optimize screening strategies and develop novel approaches to increase participation [8,9].

One potential strategy is to harness existing routinely available clinical and imaging data for individual CRC risk prediction with a particular focus on the liver. Owing to its central role in metabolic processes, the liver provides predictive potential for chronic disease risk. For example, laboratory and imaging hepatic parameters have previously been shown to predict type 2 diabetes, cardiovascular conditions and psoriasis [10-15]. In the context of CRC, the interplay within the gut–liver axis represents a critical mechanistic link: disrupted bile acid metabolism in conditions such as primary sclerosing cholangitis (PSC) and metabolic dysfunction-associated steatotic liver disease (MASLD), as well as inflammatory mediators transported via the portal system in patients with inflammatory bowel disease (IBD), contribute to hepatic and colonic pathologies [16-18]. Moreover, shared risk factors such as obesity and metabolic syndrome further link liver and colorectal diseases [11,19].

Building on these pathophysiological foundations, machine learning (ML) models provide a powerful approach to extract hepatic features from medical imaging as biomarkers for CRC risk stratification. By enabling the quantification of biomedical information, ML methods facilitate the integration of various imaging and clinical data into risk prediction algorithms, an approach that has already been successfully applied to predict e.g. cardiovascular risk based on liver magnetic resonance imaging (MRI) data [13,20].

Leveraging the gut-liver axis, this proof-of-concept study suggests that morphological features of the liver, extracted from routine CT, may serve as non-invasive markers of colorectal neoplasia as an additional opportunistic screening method besides established CRC screening methods. Importantly, this approach could make use of CT examinations performed for other clinical indications, thereby broadening the potential for early risk assessment beyond dedicated screening settings. Ultimately, this study aims to provide a first step toward low-threshold strategies for CRC screening.

## Methods

Study population

This retrospective study was approved by the local ethics committee (EK II 2023-887-AF 11). Patient data from colonoscopies performed in clinical routine between 1 January 2013 and 29 December 2023 at a single center were reviewed (n=12,355) and cross-checked with the local imaging database to determine whether an abdominal CT scan was performed within five years before or after colonoscopy, resulting in 6,331 patients. After applying further exclusion criteria (see **Figure 1**), a total of 1,997 patients were included in the final analysis. The Checklist for Artificial Intelligence in Medical Imaging (CLEAR) and Standards for Reporting of Diagnostic

Accuracy (STARD) were applied to ensure transparency, reproducibility, and high-quality reporting (**Supplementary Figures 1 and 2**) [21, 22].

Colonoscopy and liver diagnosis data

Colonoscopy, laboratory and available sonography reports examining liver diseases were reviewed by trained medical study personnel. Patients were classified according to their most advanced finding at colonoscopy as follows: i) CRC, ii) adenoma (further subclassified into advanced and non-advanced adenomas), or iii) no colorectal neoplasia. Advanced adenomas were defined by at least one of the following: size > 1 cm, tubulovillous or villous architecture, or high-grade dysplasia.

Liver diagnoses were extracted from the local hospital information system (SAP Deutschland SE & Co. KG, Walldorf, Germany) based on the 10th revision of the International Statistical Classification of Diseases and Related Health Problems (ICD-10) codes K70.- to K77.-. Diagnoses were based on imaging and/or laboratory results. Patients with diagnosis of solid liver tumor, liver abscess or metastases in the liver were excluded as detailed in **Figure 1**.

Image acquisition and segmentation

All included CT scans came from a single center using ten different CT scanners (see **Supplementary Table 1** for further details). Scans were selected if they were acquired with contrast agent (Imeron®350, Bracco Imaging Deutschland GmbH, Konstanz, Germany) in portal venous phase (70 seconds post-injection). This protocol was chosen as it is the most frequently used abdominal CT protocol in clinical routine across a wide range of indications, which is essential for ensuring broad applicability of the algorithm trained on these scans.

Liver segmentations were generated by applying MultiTalent, a model built upon the nnU-Net framework, to the 3D CT scans [23,24]. MultiTalent was trained on 13 public abdominal CT datasets (1,477 images, 47 classes). For inference, we applied the pretrained model in evaluation mode, using only its liver-specific sigmoid head to generate binary liver masks for downstream analysis without additional fine-tuning.

Feature engineering

We used the Radiomics Processing Toolkit (RPTK) [25], to generate reproducible standardized radiomics features, achieving state-of-the-art quality for reproducibility and transparency according to the CLEAR guidelines (see **Supplementary Figure S1)** [22]. All images and corresponding segmentations were resampled to an isotropic voxel spacing of 1 mm³. Image resampling was performed using third-order B-spline interpolation, while segmentations were resampled using nearest-neighbor interpolation to preserve label integrity. To reduce segmentation artifacts, a connected component analysis was applied, and only the largest connected component was retained for subsequent analysis. Radiomic features were extracted without applying any additional image transformations or mask perturbations. Intensity discretization was performed using a fixed bin width of 25 prior to feature computation. All resampling procedures were implemented using the **SimpleITK** library [26]. A total of 227 radiomic features - covering eight different feature classes (see **Supplementary Figure S3**) - were extracted using PyRadiomics [27], and subsequently normalized using z-score normalization.

In addition to the liver, the liver-surrounding region was analyzed, and radiomic features were extracted and integrated into the feature space. Afterwards, correlation filtering with a pearson correlation coefficient threshold of r=0.9 and a variation filtering with a threshold of r=0.1 was performed. Feature selection was performed using sequential selection by using a random forest classifier, resulting in a final set of 18 features that were used to predict colorectal neoplasia (**Figure 2**). These features were used to predict colorectal neoplasia based on liver-related imaging features. No feature selection was performed on the test set.

Clinical features (chronic liver disease, colon pathology, age, sex, cholecystectomy and inflammatory bowel disease) were also included to identify relevant clinical risk factors. There was no significant correlation with any clinical paramter and the presence of colorectal neoplasia (see **Supplementary Figure S4**). Therefore, none of the clinical features were retained in the final model based on missing performance gain during the iterative selection process and prediction of colorectal neoplasia was based solely on radiomic features. The heatmap (**Figure 2**) illustrates clear differentiation patterns between patients with and without colorectal neoplasia based on specific radiomics-derived texture and morphological features. Feature clustering revealed that classes such as grey level co-occurrence matrix and first-order statistics contributed most prominently to this stratification.

Model training

The dataset was randomly split into training and test set using sklearn [28] (**Figure 1**). Five model types - Random Forest, TabNet, Light Gradient Boost Model (LGBM), Extreme Gradient Boost Model (XGBoost), Support Vector Machine (SVM) - were trained, cross-validated (five-fold), and optimized on the training set, which included 1397 randomly selected cases. The best optimized ensembled model were reported and applied to the held-out test set (including 600. Cases). SMOTE (Synthetic Minority Over-sampling Technique) was applied to address class imbalance in the training dataset for the sub-classifications of CRC vs. adenoma and CRC/adenoma vs. no colorectal neoplasia. After training the five models on 5 different folds, the fold-specific models were ensembled. The best model type was selected based on its averaged cross-validation AUROC.

Statistical analysis

Model performance was evaluated using the area under the receiver operating characteristic curve (AUROC), sensitivity and specificity. The optimal classification threshold was derived by maximizing the Youden Index (sensitivity + specificity − 1) on the training dataset. This threshold was subsequently applied to the test dataset to assess model performance in an unbiased manner and to prevent information leakage. Correlations between clinical parameters were assessed using the Pearson correlation coefficient. Descriptive values are presented as means with standard deviations (SD) or as percentages (%), as appropriate. Statistical analyses were performed using RStudio (PositPBC, Boston, Massachusetts, US, Version 2024.12.1). For the confusion matrix evaluation and ROC analysis, sklearn (version 1.5.0) was used [28]. A summary of the data processing approach is given in **Figure 3**.

**Results**

Prediction of colorectal neoplasia based on liver radiomics

Multiple ML models were applied to predict the presence of colorectal neoplasia based on radiomics extracted from the liver on CT. After training on the selected radiomic features from PyRadiomics with RPTK, the best-performing model was the XGBoost model (Validation AUROC: 0.832 ± 0.013; Test AUROC: 0.810 [95% CI: 0.767 - 0.837] (**Supplementary Figure S5**). In contrast, a model based solely on clinical data performed significantly worse, with the best clinical-only XGBoost model yielding near-random classification performance (Validation AUROC: 0.547 ± 0.018; Test AUROC: 0.457 [95% CI: 0.411-0.506] (**Supplementary Figure S6**).

Model selection was based on the highest validation AUROC to ensure robust generalizability (see **Supplementary Figures S5** and **S6**). **Figures 5a and 5b** display the corresponding ROC curves for the cross-validation training and the ensembled model applied to the test set.

<u>Sensitivity and Specificity for classification of colorectal neoplasia vs. no colorectal neoplasia</u>

The evaluation of the sensitivity and specificity analysis in the context of classifying colorectal neoplasia vs. the absence of colorectal neoplasia demonstrated improved performance of the XGBoost model following threshold optimization via the Youden Index. As shown in **Figure 4e**, the initial confusion matrix (left) reflected a relatively high specificity (95.6%) but a markedly lower sensitivity (23.0%) on the test set. After weighting of sensitivity and specificity and post-adjustment of the classification threshold based on the Youden J statistic (J = 0.667), the best performing corrected model (right) achieved a more balanced performance, with a sensitivity of 74.1% and a specificity of 72.3%. The model also showed generalizability, with training set sensitivity and specificity of 87.9% and 76.7% (**Figure 4c and 4d**), respectively, and corresponding test set values of 74.1% and 72.3% (**Figure 4c and 4e**). When further analyzing the true positive and false negative predicted cohort, we found that only 14.9% ± 36.0% of the patients had liver diseases and 5.7% ± 34.0% in the cohort of false negatives had diagnosed liver diseases **(Figure 4f)**.

Radiomics-based differentiation of CRC and adenoma

To further evaluate the model's capacity for sub-classification within colorectal neoplasia, we assessed its performance in differentiating between adenoma and CRC using both radiomic and clinical features. The best-performing approach utilized radiomics-derived features in combination with clinical data, achieving an AUROC of 0.671 ± 0.018 on the validation set and AUROC 0.674 [ 95% CI: 0.606 - 0.741] on the test set using an XGBoost classifier (**Figure 5a**). This indicates moderate discriminative ability between adenoma and CRC lesions (**Supplementary Figure S7**). Further classification of adenoma, or CRC against healthy patients resulted in validation AUROC: 0.913 ± 0.014 and test AUROC: 0.659 [95% CI: 0.606-0.712] with Random Forest Classifier for CRC (**Figure 5b**, **Supplementary Figure S8**) and validation AUROC: 0.873 ± 0.009 and test AUROC: 0.651 [95% CI: 0.592-0.705] with XGBoost for adenoma (**Figure 5c, Supplementary Figure S9**).

**Discussion**

This study demonstrates that radiomics features extracted from the liver on contrast-enhanced abdominal CT scans can non-invasively predict colorectal neoplasia with high accuracy. In a large, real-world cohort of nearly 2,000 patients who underwent both colonoscopy and abdominal CT, our XGBoost model achieved robust predictive performance (validation AUROC: 0.832 ± 0.013; test AUROC: 0.810 [95% CI: 0.767–0.837]). These results indicate that liver-derived imaging features may reflect hepatic processes associated with colorectal tumorigenesis, highlighting the gut–liver axis as a relevant biomarker source. Consequently, our findings offer a foundation for new biomechanical screening tools and support integration of such tools into national screening programs in addition to established methods such as colonoscopy, particularly relevant given persistently low participation rates in CRC screening in Germany and other countries.

The high number of CT scans conducted annually in Germany, estimated at 160 per 1,000 individuals or approximately 13.3 million scans in 2021 [29], underlines the potential public health impact of such an opprotunistic screening approach. If colorectal neoplasia is predicted, this information could be directly communicated in radiology reports, thereby prompting timely recommendations for colonoscopy. This approach addresses key barriers to traditional screening, such as patient discomfort and low participation, by intervening in the clinical workflow where patients are already engaged [30,31].

Our method may also enhance post-screening surveillance. Post-imaging colorectal cancer (PICRC) remains a challenge, with rates of 4–6% [32-34]; our strategy could offer early identification of patients at risk for PICRC in the interval between colonoscopies.

While the association between chronic liver disease and CRC is well known [11,17,18], automated colorectal neoplasia prediction based on biomedical features extracted from the liver is not established in organized screening programs. A comparable approach, using radiomics features from a surrogate or remote organ has already been validated, for example, using pancreatic volume to predict cancer cachexia in head and neck cancer patients [35], and for the prediction of multiple chronic diseases based on liver features [12-15], highlighting the potential of utilizing inter-organ crosstalk for cancer screening, therapy monitoring, and advanced risk stratification.

In line with more established screening modalities like FOBT [36], differentiation between advanced neoplastic subtypes remains moderate in our study with an AUROC for adenoma detection of 0.651 [95% CI: 0.592–0.705] and an AUROC for distinguishing adenomas from CRC of 0.674 [95% CI: 0.606–0.741], likely reflecting subtler changes in early adenomatous stages and class imbalance.

Previous studies evaluating clinical risk models have reported moderate diagnostic performance with an exemplary AUROC of 0.75 by Chen et al. [37], in our study, clinical data alone even showed poor performance (test AUROC: 0.457 [95% CI: 0.411–0.506]) and was substantially worse than the radiomics-based approach (test AUROC: 0.810 [95% CI: 0.767–0.837]). Our performance on the clinical data was considerably lower compared to Chen et al., which can be attributed to the limited number of clinical parameters included as well as the very homogeneous age and sex distribution between patients with and without colorectal neoplasia.

Nevertheless, these results illustrate that radiomics signatures may capture subtle, even subclinical, hepatic changes linked to colorectal neoplasia, as suggested by the finding that only 14.9% ± 36.0% of correctly classified positive cases had documented liver disease.

The broader field has already demonstrated high diagnostic accuracy of radiomic models for other malignancies such as hepatocellular carcinoma or pancreatic ductal adenocarcinoma, reporting AUROCs above 0.90 [38,39]. Adoption of radiomics for CRC screening has been limited, likely due to technical challenges in colon segmentation [40]; therefore, our solution focuses on the liver as a surrogate, which may facilitate wider implementation and further validation with direct analysis on segmented colon abnormalities.

While our model's sensitivity (74.1%) and specificity (72.3%) are encouraging, they remain moderate when considering the requirements for a primary screening tool [41]. In general, it is important to emphasize that we do not propose replacing established CRC screening methods, but rather aim to provide a complementary approach, particularly for patients with existing CT data, where an opportunistic strategy can be applied. To further improve future opportunistic screening approaches, integration of liver radiomics with other non-invasive risk factors such as liquid biopsy [42], and advanced imaging modalities (e.g. MRI or ultrasound-based elastography) [43] as well as inclusion of additional clinical or lifestyle data could be considered.

Key limitations of our study include its retrospective, single-center design and the heterogeneity of indications for colonoscopy, which may introduce selection bias. The temporal sequence, where some CT scans were performed after adenoma removal, may attenuate detected associations. Moreover, our cohort shows a high prevalence of CRC (48%) compared to the prevalence of adenomas (52%), which reflects the university hospital setting where many patients are referred for carcinoma treatment, while only a small proportion undergo initial screening. External validation across larger, prospectively recruited, multicenter screening cohorts is essential before clinical implementation.

In summary, AI-based analysis of liver-derived radiomic features from routine CT represents a promising and widely accessible clinical tool for CRC risk stratification. Embedding this approach into established clinical workflows could enhance screening uptake and facilitate earlier CRC detection and prevention. Ultimately, by exploring the the gut-liver axis, this strategy may spark new mechanistic hypotheses that shape the next generation of precision prevention strategies.

monitoring in colorectal cancer. Mol Cancer. 2022 Mar 25;21(1):86. doi: 10.1186/s12943-022-01556-2. PMID: 35337361; PMCID: PMC8951719.
43. Maino C, Vernuccio F, Cannella R, Franco PN, Giannini V, Dezio M, Pisani AR, Blandino AA, Faletti R, De Bernardi E, Ippolito D, Gatti M, Inchingolo R. Radiomics and liver: Where we are and where we are headed? Eur J Radiol. 2024 Feb;171:111297. doi: 10.1016/j.ejrad.2024.111297. Epub 2024 Jan 12. PMID: 38237517.

# Figures and Tables

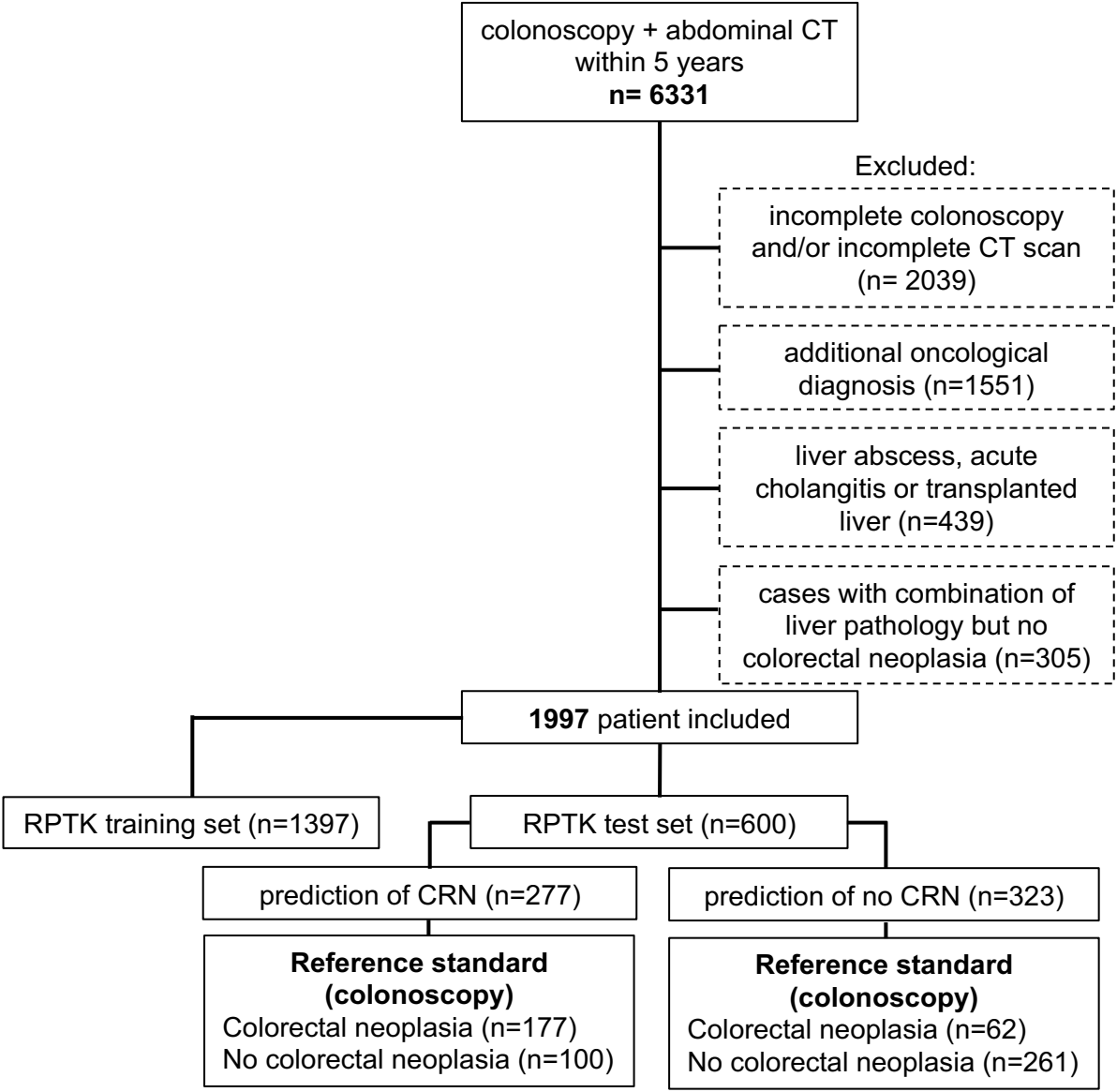

*Figure 1.* STARD flow-chart of patients' cohort, the training set and test set. Additional solid tumors refer to all oncological diagnoses of solid cancers that have a known tendency of metastasizing into the liver (such as breast cancer) and solid cancer with the liver being the known primary region (e.g. hepatocellular carcinoma). Colorectal neoplasia was diagnosed and characterized during colonoscopy in clinical routine. Liver diagnoses were extracted from the internal clinical reporting system and are based on imaging and/or laboratory diagnosis. Incomplete colonoscopy is defined as colonoscopy which did not reach the coecum. RPTK: Radiomics Processing ToolKit; CRN: colorectal neoplasia.

| Patients' characteristics | n (%) |
|---|---|
| men | 1138 (57%) |
| women | 859 (43%) |
| mean age at CT scan (years) | 64 |
| mean time between colonoscopy and CT (months) | 7 |
| CT prior colonoscopy | 1650 (83%) |
| IBD | 158 (8%) |
| CHE | 207 (1%) |
| colorectal neoplasia | 808 (40%) |
|    CRC | 385 (48%) |
|    adenoma | 423 (52%) |
|       advanced adenoma | 141 (33%) |
|    right sided neoplasia | 336 (41%) |
|    no liver pathology | 643 (79%) |
|    diagnosed liver pathology | 165 (21%) |
|       MASLD | 74 (44%) |
|       toxic and alcoholic liver diseases | 15 (8%) |
|       cirrhosis | 76 (47%) |
| no colorectal neoplasia or liver pathology | 1189 (60%) |

***Table 1.*** *Patients' characteristics. IBD: inflammatory bowel disease; CHE: cholecystectomy; CRC: colorectal cancer; MASLD: metabolic associated steatotic liver disease.*

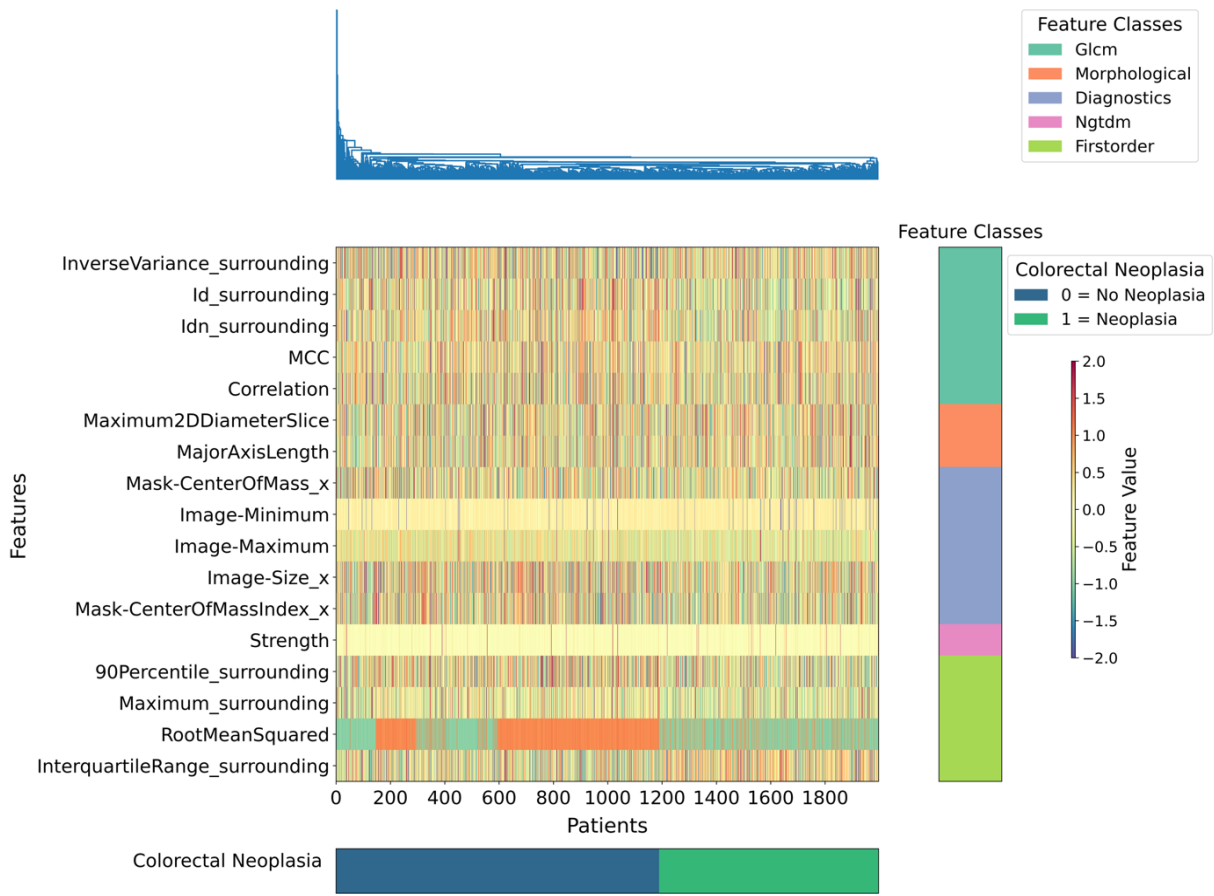

*Figure 2.* Stratification of selected radiomics features from PyRadiomics for colorectal neoplasia based on the features extracted from the liver with a best straifying root mean squared and 90th percentile. No clinical parameter was determined to be relevant (Id: Inverse Difference, Idn: Inverse Difference Normalized, MCC: Maximal Correlation Coefficient).

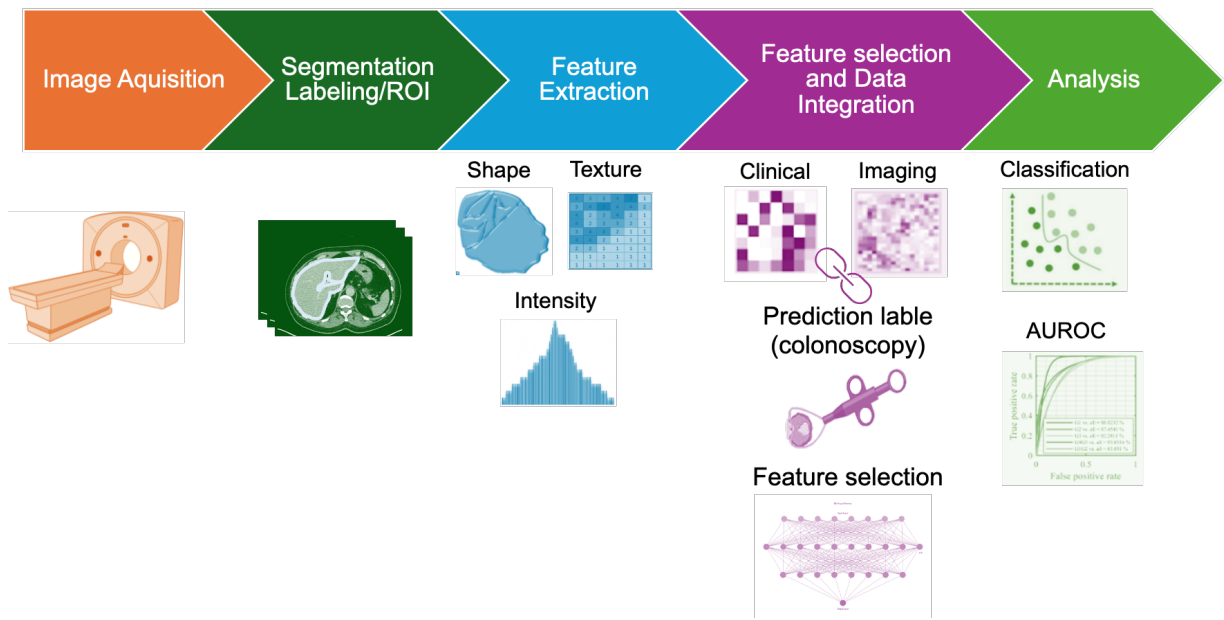

*Figure 3.* Data processing flow-chart. The workflow of LIRA includes five continuous parts: (1) Image acquisition of abdominal CT scans. (2) Liver segmentation, with automatic segmentation methods. (3) Feature extraction. (4) Model building and data integration of colonoscopy diagnosis functioning as prediction target. (5) Statistical analysis of model performance. Abbreviations: AUROC: area under the receiver operating curve. Own illustration.

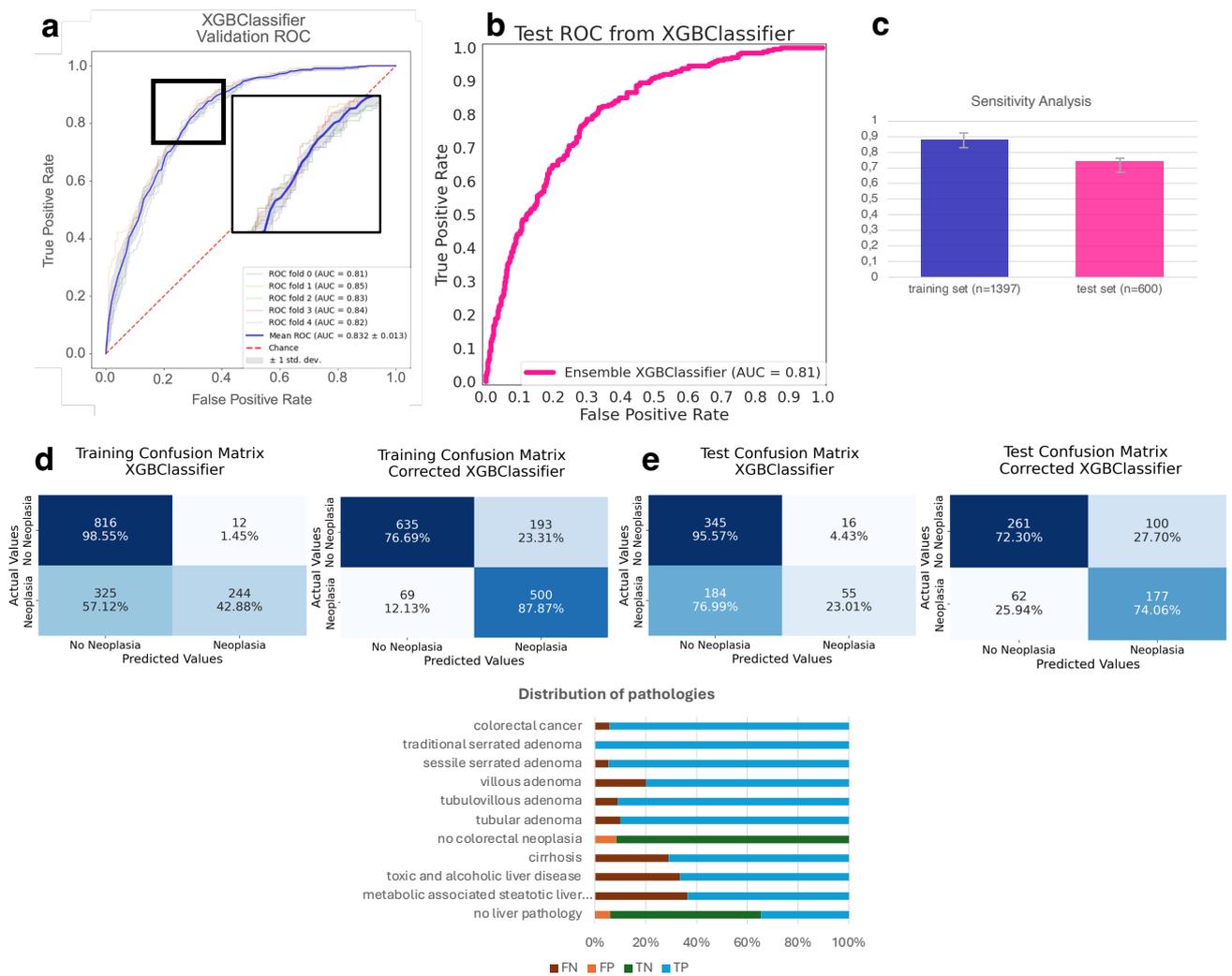

*Figure 4. **a**, ROC Curves of the 5-fold validation set of the best-performed model (XGBoost Classifier) for colorectal neoplasia prediction. **b**, ROC curve ensemble on the test set of the best-performed model (XGBoost Classifier) for colorectal neoplasia prediction.* **c**, sensitivity of *XGBoost Classifier for colorectal neoplasia prediction* in the trainings set (n=1,397) and test set (n=600). The error bars denote the two-sided 95% CI. **d**, c*onfusion matrices of the best model on the training set before (left) and after (right) Youden index correction for optimization of threshold-based matrices.* **e**, c*onfusion matrices of the best model on the test set before (left) and after (right) Youden index correction for optimization of threshold-based matrices (Label: 0= no colorectal neoplasia, 1= colorectal neoplasia).* **f**, *Disruption of liver and colon pathologies within the test set, subclassified in true positive (TP), false negative (FN), false positive (FP) and true negative (TN) results for the prediction of colorectal neoplasia of the XGBoost Classifier.*

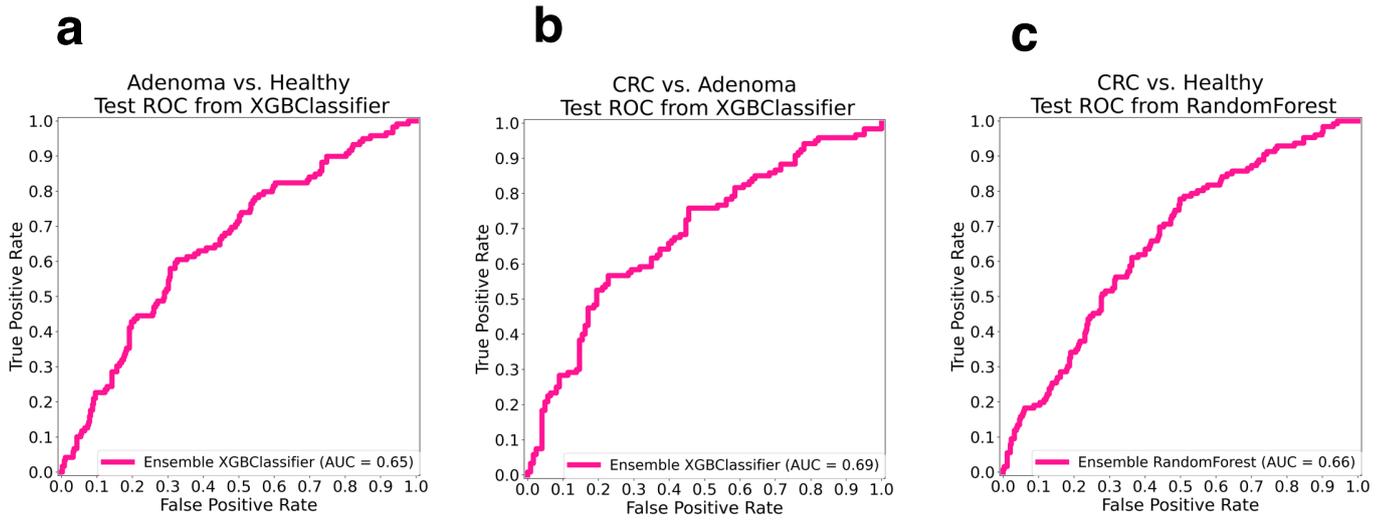

*Figure 5. **a**, ROC curve ensemble on the test set of the best-performed model (XGBoost Classifier) for discrimination of CRC and adenoma. **b**, ROC curve ensemble on the test set of the best-performed model (*Random Forest Classifier*) for discrimination of CRC and healthy patients. **c**, ROC curve ensemble on the test set of the best-performed model (XGBoost Classifier) for discrimination of adenoma and healthy patients.*

# Supplementary Figures and Tables

| Section & Topic | No | Item | |
|---|---|---|---|
| **TITLE OR ABSTRACT** | | | |
| | 1 | Identification as a study of diagnostic accuracy using at least one measure of accuracy (such as sensitivity, specificity, predictive values, or AUC) | 5-6 |
| **ABSTRACT** | | | |
| | 2 | Structured summary of study design, methods, results, and conclusions (for specific guidance, see STARD for Abstracts) | 5-6 |
| **INTRODUCTION** | | | |
| | 3 | Scientific and clinical background, including the intended use and clinical role of the index test | 7-8 |
| | 4 | Study objectives and hypotheses | 8 |
| **METHODS** | | | |
| *Study design* | 5 | Whether data collection was planned before the index test and reference standard were performed (prospective study) or after (retrospective study) | 8 |
| *Participants* | 6 | Eligibility criteria | 8-9 |
| | 7 | On what basis potentially eligible participants were identified (such as symptoms, results from previous tests, inclusion in registry) | 8-9 |
| | 8 | Where and when potentially eligible participants were identified (setting, location and dates) | 8-9 |
| | 9 | Whether participants formed a consecutive, random or convenience series | 11 |
| *Test methods* | 10a | Index test, in sufficient detail to allow replication | 9-11 |
| | 10b | Reference standard, in sufficient detail to allow replication | 9 |
| | 11 | Rationale for choosing the reference standard (if alternatives exist) | 7,9 |
| | 12a | Definition of and rationale for test positivity cut-offs or result categories of the index test, distinguishing pre-specified from exploratory | 11-12 |
| | 12b | Definition of and rationale for test positivity cut-offs or result categories of the reference standard, distinguishing pre-specified from exploratory | 11-12 |
| | 13a | Whether clinical information and reference standard results were available to the performers/readers of the index test | n/a |
| | 13b | Whether clinical information and index test results were available to the assessors of the reference standard | n/a |
| *Analysis* | 14 | Methods for estimating or comparing measures of diagnostic accuracy | 11-12 |
| | 15 | How indeterminate index test or reference standard results were handled | n/a |
| | 16 | How missing data on the index test and reference standard were handled | n/a |
| | 17 | Any analyses of variability in diagnostic accuracy, distinguishing pre-specified from exploratory | 10-12 |
| | 18 | Intended sample size and how it was determined | n/a |
| **RESULTS** | | | |
| *Participants* | 19 | Flow of participants, using a diagram | Figure 1 |
| | 20 | Baseline demographic and clinical characteristics of participants | Table 1 |
| | 21a | Distribution of severity of disease in those with the target condition | Tabel 1 |
| | 21b | Distribution of alternative diagnoses in those without the target condition | Tabel 1 |
| | 22 | Time interval and any clinical interventions between index test and reference standard | Tabel 1 |
| *Test results* | 23 | Cross tabulation of the index test results (or their distribution) by the results of the reference standard | Figure 5 |
| | 24 | Estimates of diagnostic accuracy and their precision (such as 95% confidence intervals) | 12-13 |
| | 25 | Any adverse events from performing the index test or the reference standard | n/a |
| **DISCUSSION** | | | |
| | 26 | Study limitations, including sources of potential bias, statistical uncertainty, and generalisability | 17 |
| | 27 | Implications for practice, including the intended use and clinical role of the index test | 17 |
| **OTHER INFORMATION** | | | |
| | 28 | Registration number and name of registry | n/a |
| | 29 | Where the full study protocol can be accessed | 1 |
| | 30 | Sources of funding and other support; role of funders | 1 |

*Figure S1. STARD Checklist [24]; n/a: not applicable.*

# CLEAR Checklist v1.0

**Note**: Use the checklist in conjunction with the main text for clarification of all items.
Yes, details provided; No, details not provided; n/e, not essential; n/a, not applicable; Page, page number

| Section | No. | Item | Yes | No | n/a | Page |
|---|---|---|---|---|---|---|
| **Title** | | | | | | |
| | 1 | Relevant title, specifying the radiomic methodology | | ☐ | ☐ | 0 |
| **Abstract** | | | | | | |
| | 2 | Structured summary with relevant information | | ☐ | ☐ | 5-6 |
| **Keywords** | | | | | | |
| | 3 | Relevant keywords for radiomics | | ☐ | ☐ | 6 |
| **Introduction** | | | | | | |
| | 4 | Scientific or clinical background | | ☐ | ☐ | 7-8 |
| | 5 | Rationale for using a radiomic approach | | ☐ | ☐ | 7-8 |
| | 6 | Study objective(s) | | ☐ | ☐ | 8 |
| **Method** | | | | | | |
| *Study design* | 7 | Adherence to guidelines or checklists (e.g., CLEAR checklist) | | ☐ | ☐ | 8 |
| | 8 | Ethical details (e.g., approval, consent, data protection) | | ☐ | ☐ | 8 |
| | 9 | Sample size calculation | ☐ | ☐ | | |
| | 10 | Study nature (e.g., retrospective, prospective) | | ☐ | ☐ | 8 |
| | 11 | Eligibility criteria | | ☐ | ☐ | 9 |
| | 12 | Flowchart for technical pipeline | | ☐ | ☐ | 25 |
| *Data* | 13 | Data source (e.g., private, public) | | ☐ | ☐ | 8 |
| | 14 | Data overlap | ☐ | ☐ | | |
| | 15 | Data split methodology | | ☐ | ☐ | 11 |
| | 16 | Imaging protocol (i.e., image acquisition and processing) | | ☐ | ☐ | 9-10 |
| | 17 | Definition of non-radiomic predictor variables | | ☐ | ☐ | 9 |
| | 18 | Definition of the reference standard (i.e., outcome variable) | | ☐ | ☐ | 9 |
| *Segmentation* | 19 | Segmentation strategy | | ☐ | ☐ | 9-10 |
| | 20 | Details of operators performing segmentation | | ☐ | ☐ | 9-10 |
| *Pre-processing* | 21 | Image pre-processing details | | ☐ | ☐ | 9-10 |
| | 22 | Resampling method and its parameters | ☐ | ☐ | | |
| | 23 | Discretization method and its parameters | | ☐ | ☐ | 9-10 |
| | 24 | Image types (e.g., original, filtered, transformed) | | ☐ | ☐ | 9-10 |
| *Feature extraction* | 25 | Feature extraction method | | ☐ | ☐ | 10-11 |
| | 26 | Feature classes | | ☐ | ☐ | 10-11 |
| | 27 | Number of features | | ☐ | ☐ | 10 |
| | 28 | Default configuration statement for remaining parameters | ☐ | ☐ | | |
| *Data preparation* | 29 | Handling of missing data | ☐ | ☐ | | |
| | 30 | Details of class imbalance | | ☐ | ☐ | 11 |
| | 31 | Details of segmentation reliability analysis | ☐ | | ☐ | |
| | 32 | Feature scaling details (e.g., normalization, standardization) | | ☐ | ☐ | 10-11 |
| | 33 | Dimension reduction details | | ☐ | ☐ | 10-11 |
| *Modeling* | 34 | Algorithm details | | ☐ | ☐ | 10-11 |
| | 35 | Training and tuning details | | ☐ | ☐ | 11 |
| | 36 | Handling of confounders | | ☐ | ☐ | 10-11 |
| | 37 | Model selection strategy | | ☐ | ☐ | 11 |
| *Evaluation* | 38 | Testing technique (e.g., internal, external) | | ☐ | ☐ | 11 |
| | 39 | Performance metrics and rationale for choosing | | ☐ | ☐ | 11-12 |
| | 40 | Uncertainty evaluation and measures (e.g., confidence intervals) | | ☐ | ☐ | 12-14 |
| | 41 | Statistical performance comparison (e.g., DeLong's test) | | ☐ | ☐ | 11-12 |
| | 42 | Comparison with non-radiomic and combined methods | | ☐ | ☐ | 11-12 |
| | 43 | Interpretability and explainability methods | | ☐ | ☐ | 11-12 |
| **Results** | | | | | | |
| | 44 | Baseline demographic and clinical characteristics | | ☐ | ☐ | T1 |
| | 45 | Flowchart for eligibility criteria | | ☐ | ☐ | F1 |
| | 46 | Feature statistics (e.g., reproducibility, feature selection) | | ☐ | ☐ | F2 |
| | 47 | Model performance evaluation | | ☐ | ☐ | F5 |
| | 48 | Comparison with non-radiomic and combined approaches | | ☐ | ☐ | 12 |
| **Discussion** | | | | | | |

| | 49 | Overview of important findings | ☐ | ☐ | 14 |
| --- | --- | --- | --- | --- | --- |
| | 50 | Previous works with differences from the current study | ☐ | ☐ | 15, 16 |
| | 51 | Practical implications | ☐ | ☐ | 17 |
| | 52 | Strengths and limitations (e.g., bias and generalizability issues) | ☐ | ☐ | 14, 17 |
| **Open Science** | | | | | |
| *Data availability* | 53 | Sharing images along with segmentation data [n/e] | ☐ | ☐ | 1 |
| | 54 | Sharing radiomic feature data | ☐ | ☐ | 1 |
| *Code availability* | 55 | Sharing pre-processing scripts or settings | ☐ | ☐ | 1 |
| | 56 | Sharing source code for modeling | ☐ | ☐ | 1 |
| *Model availability* | 57 | Sharing final model files | ☐ | ☐ | 1 |
| | 58 | Sharing a ready-to-use system [n/e] | ☐ | ☐ | 1 |

*Figure S1. CLEAR Checklist [22].*

| | |
|---|---|
| Included kernels | B30f, B30s, B30, B31f, B40f, Br 36, Br38, Br39, Br40, Bv45, I30, I30s, I30f, I31f, Q30f |
| CT scanners used in training set | Biograph 40 (n=48; 3%), Emotion 16 (n=388; 28%), NAEOTOM Alpha (n=20; 1%), Sensation 64 (n=273; 20%), SOMATOM Definition (n=8; 1%), SOMATOM Definition AS+ (n=271; 19%), SOMATOM Definition Flash (n=295; 21%), SOMATOM Force (n=32; 2%), SOMATOM go.Up (n=63; 5%) |
| CT scanners used in test set | Biograph 40 (n=19; 3%), Emotion 16 (n=147; 25%), NAEOTOM Alpha (n=90; 3%), Sensation 64 (n=115; 19%), SOMATOM Definition (n=4; 1%), SOMATOM Definition AS+ (n=119; 20%), SOMATOM Definition Flash (n=130; 22%), SOMATOM Force (n=13; 2%), SOMATOM go.Up (n=32; 5%); Volume Zoom (n=4; 1%) |
| Slice thickness [mm] | Mean=4.6 (SD=0.80) |
| Number of slices | Mean=111,27 (SD=53,26) |
| ROI size [Voxels] | Mean=602858.49 (SD=441908) |
| Number of connected components | Mean=1.43 (SD=2.14) |
| Number of bins (with bin width of 25) | Mean=12.89 (SD=6.53) |
| Number of grey values in ROI | Mean=251.42 (SD=56.92) |
| Mean grey value in ROI | Mean=91.15 (SD=22.22) |
| Surface wavelet | Mean=44.84 (SD=11.08) |

| Mesh curvature | Mean=-0.017 (SD=0.008) |
|---|---|
| Fractal dimension 3D | Mean=2.44 (SD=0.09) |

*Table S1. Detailed information of CT values.*

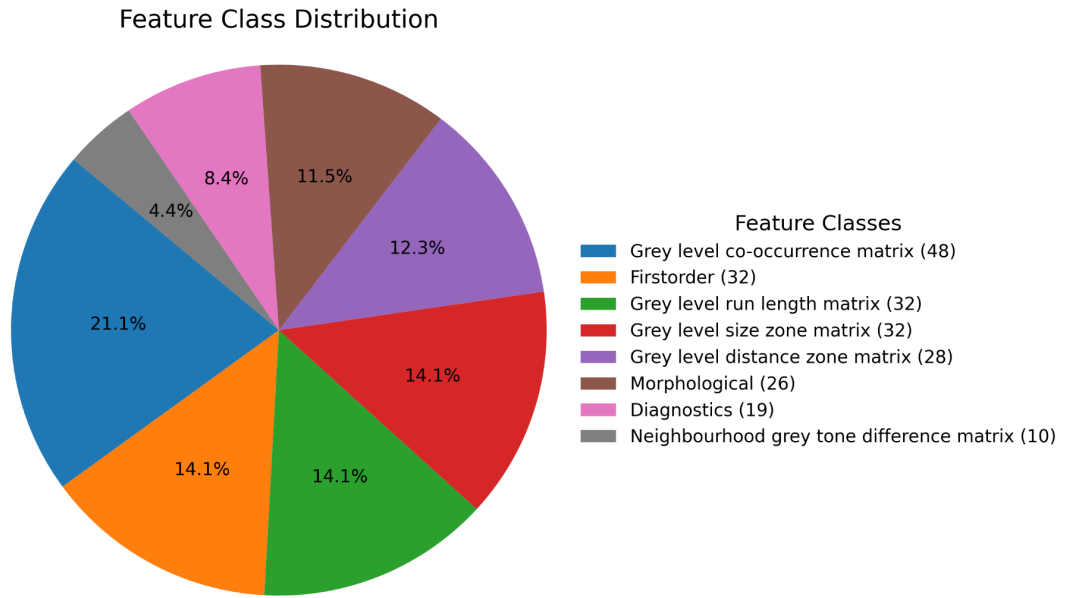

*Figure S3. Distribution of radiomics feature classes in the raw feature extraction.*

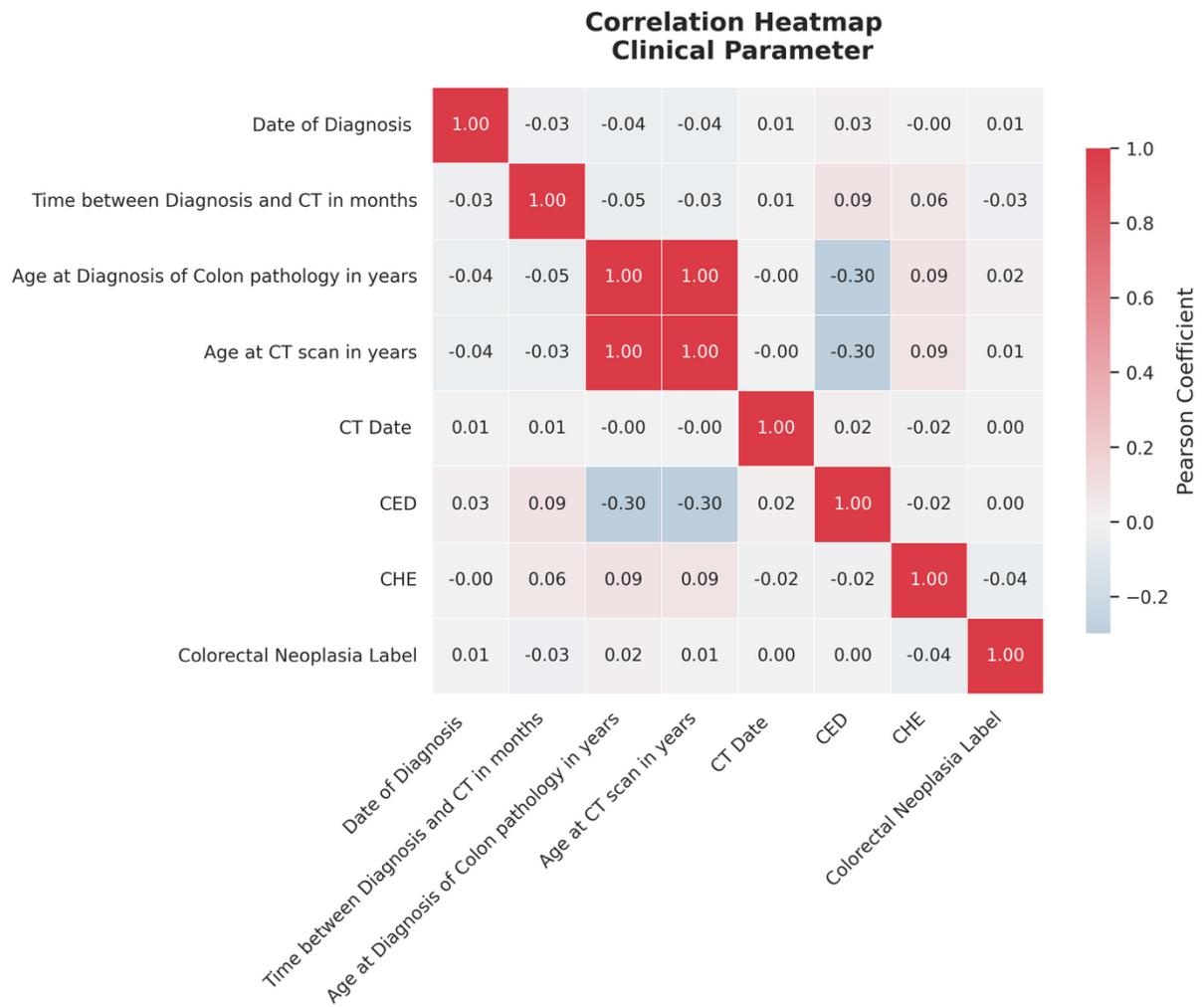

*Figure S4. Clinical feature correlation map including Pearson correlation coefficient to the colorectal neoplasia.*

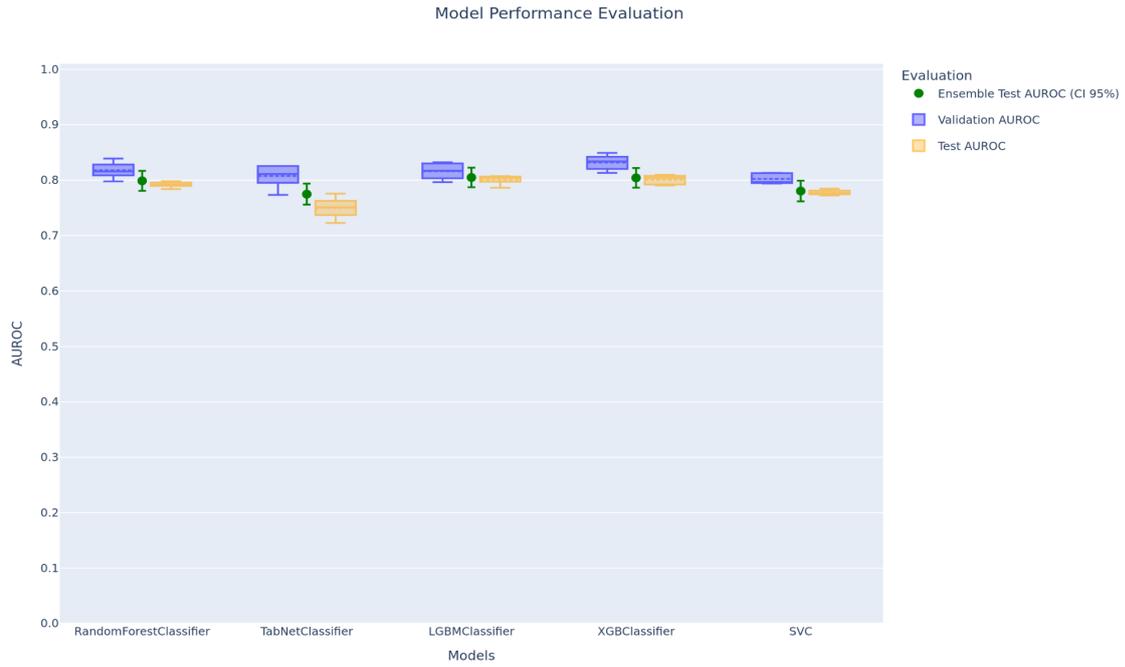

*Figure S5. Summary of AUROC performance of applied models on PyRadiomics features for colorectal neoplasia prediction using the RPTK framework.*

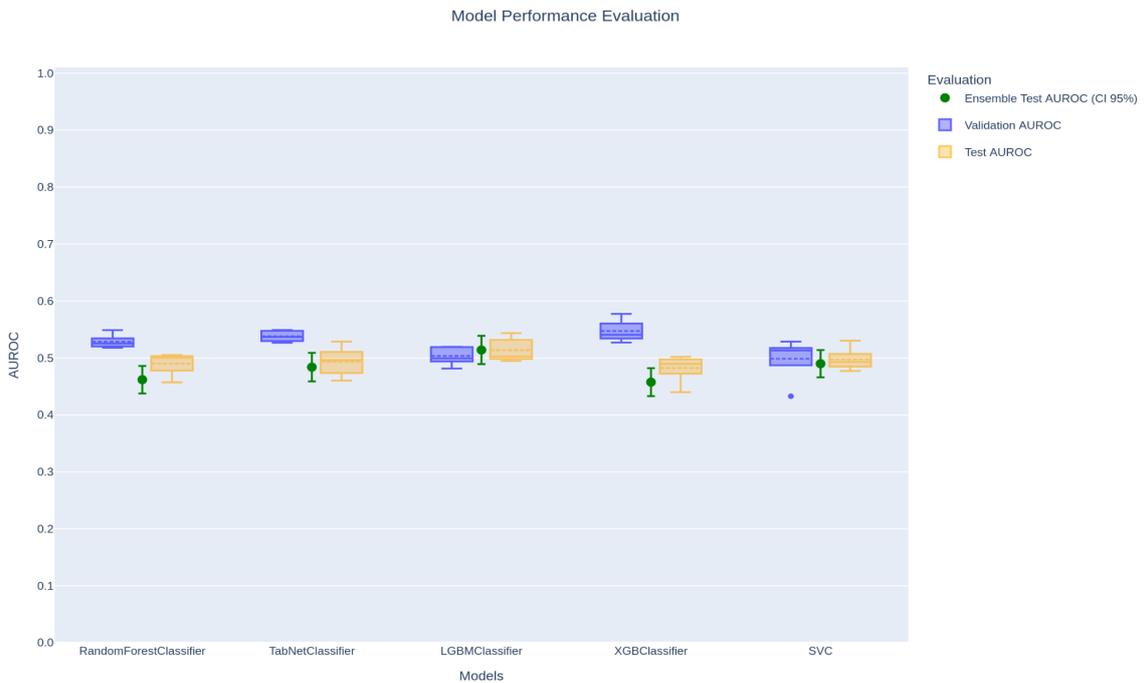

*t*

*Figure S6: Summary of AUROC performance of applied models on clinical features for colorectal neoplasia prediction using*

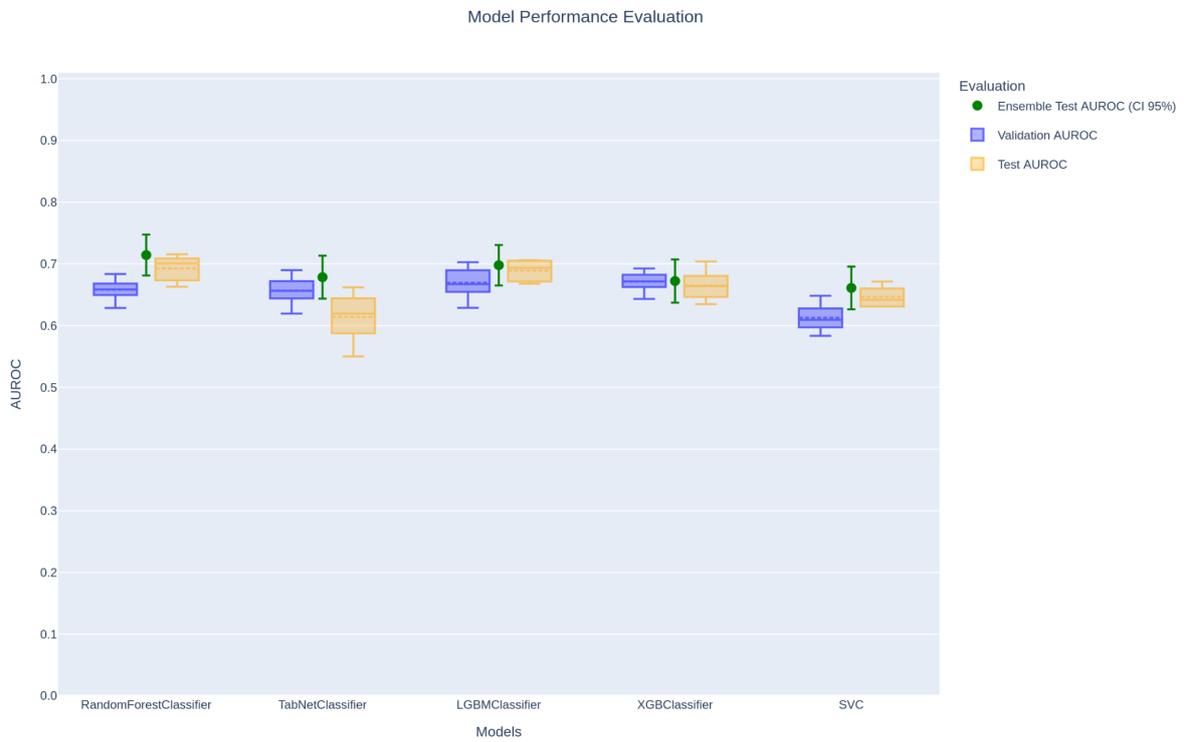

*Figure S7. Model performance for adenoma vs. CRC classification based on PyRadiomics features using the RPTK framework.*

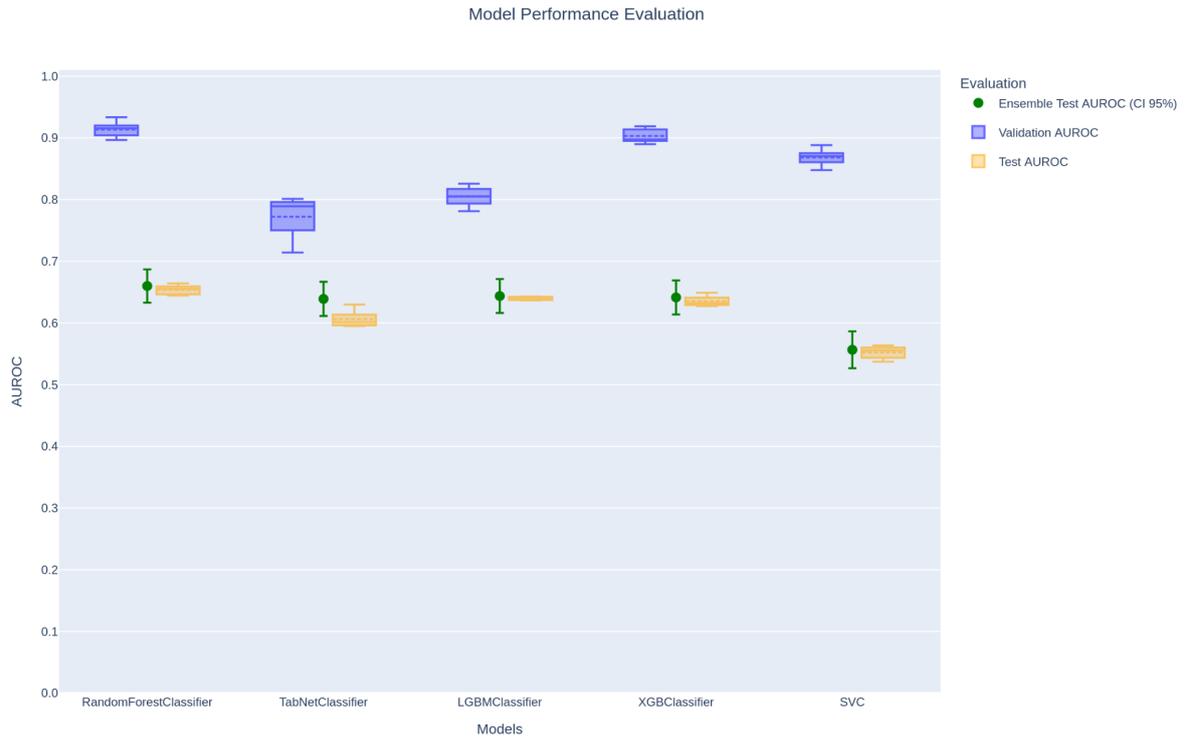

*Figure S8. Model performance for CRC vs. no colorectal neoplasia classification based on PyRadiomics features using the RPTK framework.*

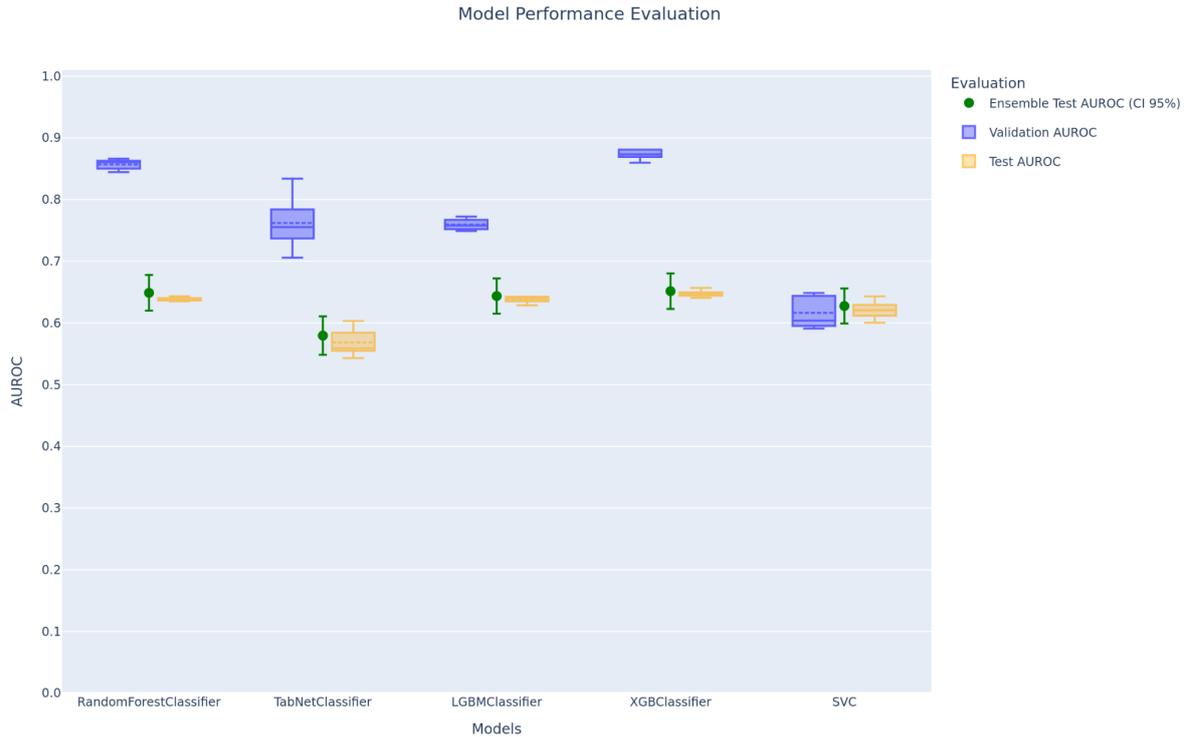

*Figure S9. Model performance for adenoma vs. no colorectal neoplasia classification based on PyRadiomics features using the RPTK framework.*

## Author Contributions

Conceptualization: A.H., F.G., J.B., S.G.; Data curation: A.H., J.B., D.T., T.N., F.I., S.B., J.B.; Formal analysis: A.H., J.B.; Investigation: A.H., J.B. N.K., J.D.; Methodology: A.H., J.B.; Project administration: A.H., F.G., J.B.; Resources: S.B., J.B., S.S., S.L., R.F.; Software: J.B., D.T., T.N., F.I., K.MH.; Supervision: F.G., K. MH.; Validation: F.G., J.B., K. MH.; Writing – original draft: A.H., J.B., D.T.; Writing – review & editing: F.G., J.B., J.N.K.

## Data transparency statement

Data, analytic methods, and study materials are available upon reasonable request.

## Code availability statement

The RPTK tool used for radiomics experiments in this study is publicly available on GitHub: https://github.com/MIC-DKFZ/RPTK

## Abbreviations

| | |
|---|---|
| AUROC | Area Under the Receiver Operating Characteristic Curve |
| CLEAR | Checklist for Artificial Intelligence in Medical Imaging |
| CTC | Computed Tomographic Colonography |
| CRC | Colorectal Cancer |
| FOBT | Fecal Occult Blood Test |
| gFOBT | Guaiac-based Fecal Occult Blood Test |

| | | |
|---|---|---|
| iFOBT | Immunochemical Fecal Occult Blood Test | |
| IBD | Inflammatory Bowel Disease | |
| ICD-10 | International Statistical Classification of Diseases and Related Healthn Problems, 10th Revision | |
| J | Youden-Index | |
| LGBM | Light Gradient Boosting Model | |
| MASLD | Metabolic Dysfunction-Associated Steatotic Liver Disease | |
| MRI | Magnetic Resonance Imaging | |
| nnUNet | No-New-Net (automatisiertes Deep-Learning-Segmentierungsmodell) | |
| OECD | Organisation for Economic Co-operation and Development | |
| PICRC | Post-Imaging Colorectal Cancer | |
| PSC | Primary Sclerosing Cholangitis | |
| RPTK | Radiomics Processing ToolKit | |
| SD | Standard Deviation | |
| SMOTE | Synthetic Minority Over-sampling Technique | |
| SAP | Systeme, Anwendungen und Produkte in der Datenverarbeitung | |
| SVM | Support Vector Machine | |
| XGBoost | Extreme Gradient Boosting | |